\pdfoutput=1
\documentclass[twocolumn,showpacs,preprintnumbers,amsmath,amssymb]{revtex4}
\usepackage{amssymb}
\usepackage{amsmath}
\usepackage{graphicx}
\usepackage{bm}
\usepackage{amsfonts}

\newcommand{\ket}[1]{\mbox{$|#1\rangle$}}
\newcommand{\bra}[1]{\mbox{$\langle#1|$}}

\def\be{\begin{equation}} 
\def\ee{\end{equation}}
\def\gsim{\mathrel{\rlap{\lower2pt\hbox{\hskip1pt\scriptsize $\sim$}}
    \raise2pt\hbox{\scriptsize $>$}}}              
\def\lsim{\mathrel{\rlap{\lower2pt\hbox{\hskip1pt\scriptsize $\sim$}}
    \raise2pt\hbox{\scriptsize $<$}}}              

\begin{document}

\title
{Decoherence of floating qubits due to capacitive coupling}

\author{Matthias Steffen}
\email{msteffe@us.ibm.com}
\author{Frederico Brito}
\author{David DiVincenzo}
\author{Shwetank Kumar}
\author{Mark Ketchen}

\affiliation
{IBM Watson Research Center, Yorktown Heights, NY 10598}



\begin{abstract}
It has often been assumed that electrically floating qubits, such as flux qubits, are immune to decoherence due to capacitive coupling.  We show that capacitive coupling to bias leads can be a dominant source of dissipation, and therefore of decoherence, for such floating qubits.  Classical electrostatic arguments are sufficient to get a good estimate of this source of relaxation for standard superconducting qubit designs. We show that relaxation times can be improved by designing floating qubits so they couple symmetrically to the bias leads. Observed coherence times of flux qubits with varying degrees of symmetry qualitatively support our results.
\end{abstract}

\volumeyear{year}
\volumenumber{number}
\issuenumber{number}
\eid{identifier}
\date{\today}

\maketitle

\section{Introduction}
A quantum computer architecture based on superconducting thin film wires and Josephson junctions is attractive in large part because of its compatibility with current state of the art fabrication methods for solid state devices. A wide array of impressive experimental demonstrations of this include single and two-qubit gates using a variety of superconducting qubits \cite{Vion02,Wallraff05,Chiorescu03,Steffen06a,Yamamoto03,Steffen06b,Sillanpaa07,Majer07,Plantenberg07}. In parallel, coherence times of qubits have increased from tens of nanoseconds to a few microseconds through an improved understanding of microwave engineering \cite{Bertet05,Houck08} and materials research \cite{Martinis05,Neeley08}. Nonetheless, there is still much to be learned about coherence times in these systems. While some qubits appear limited by two-level defects \cite{Neeley08,Houck08}, a clear explanation of the decoherence processes that affect many of these qubits remains elusive \cite{Plantenberg07,Steffen07unpub}.

Here, we analyze a previously overlooked dissipation (and therefore decoherence) channel that could explain some of the observed short coherence times. We will show that capacitive coupling to bias leads can be a significant source of relaxation in the form of spontaneous emission via electric dipole transitions even for floating flux qubits. Such dissipation mechanism has been well understood for charge qubits \cite{Ithier05} but it has, to our best knowledge, not yet been considered for flux qubits. When flux qubits are floating it has been assumed that the connection of the qubit to ground is poor and therefore such qubits are immune to capacitive coupling. This assumption is shown to be incorrect. The reactance $Y=i\omega C_g$ of the capacitance to ground $C_g$, even for floating qubits, becomes sizeable for frequencies $\omega$ in the microwave range. Thus, the coupling via this capacitance to (resistive) bias leads becomes important as our estimates show below. As we know from the formulas for the capacitance to ground in simple geometries such as discs and loops, the scale of $C_g$ is basically fixed by the overall physical size of the qubit device, and cannot be much altered by details of device shape or geometry.

The coherence of a superconducting qubit is obviously a quantum-mechanical phenomenon.  But this paper will present relaxation-time estimates based purely on {\em classical} electric circuit theory applied to linear (RLC) circuits.  Of course, Josephson junction devices can be strongly non-linear; but in cases of current interest experimentally, they operate in a nearly linear regime, in which their functioning in a circuit can be modeled by a simple inductor.  Our classical calculations then assume the basic form of an $RC$ time constant, which makes it easy to gain an intuitive understanding of the features that determine the short relaxation times of these qubits. At the same time, these classical calculations are very informative about the quantum behavior of these devices, because of the limit that the relaxation time $T_1$ puts on the time $T_2$ for the decay of quantum coherence: $T_2\leq 2 T_1$.  Previous, fully quantum mechanical calculations \cite{Burkard04} confirm that the small anharmonicities present in our system do not strongly change the computed values of $T_1$ \cite{commentc}.

Our modeling shows that the strong relaxation due to this capacitive coupling can be mitigated by {\em symmetrically} coupling the qubit to the bias leads and by engineering the admittance of the bias leads. The classical physics makes it very clear that by configuring the bias leads in such a way that no circulating currents can be generated in the qubit, relaxation times will significantly increase. The fact that symmetry is crucial is qualitatively supported by various experimental results: Symmetric qubits perform better than asymmetric ones, and we believe the reason for this involves capacitive coupling.  Naturally, a long $T_1$ does not guarantee a long $T_2$, and our calculations here will not address the many other mechanisms that are being investigated for the loss of quantum phase coherence. But since long coherence times are possible only in systems with long relaxation time, our calculations show a necessary set of conditions for achieving high quantum coherence.

Our models focus on simulating a RF SQUID - the simplest implementation of a floating flux qubit \cite{Friedman00}. The simulations can be extended to other qubit designs with multiple Josephson junctions, and we believe that the underlying arguments for decoherence will not be drastically altered.

Our calculations indicate that there is not such a large difference between floating qubits and "grounded qubits" -- ones connected to electrical ground via a direct metallic contact. It has been previously understood \cite{Oconnell08,Houck08,Martinis03} that when qubits are grounded, they are susceptible to decoherence via capacitive coupling to bias leads. Phase qubits, for example, are excited via a small coupling capacitor and therefore decohere via the same mechanism, although the coupling capacitance is generally small enough not to have an impact on current coherence times \cite{Steffen06b}. But we show here that most floating qubits in fact also have a strong (reactive) coupling to ground, because the capacitance to ground of an isolated object scales only with its linear dimension instead of volume or area and is thus appreciable for all but the smallest of flux qubit designs.

\section{Circuit Modeling}
The scale of the capacitance to ground of a flux qubit can be estimated by using several well-known results: The capacitance of a sphere of diameter $D$ to a ground at infinity is $C_{g,sphere}=2\pi\epsilon_0D$ \cite{Purcell85}. The capacitance of a disc with diameter $D$ is $C_{g,disc}=4\epsilon_0D$, differing only by the factor $2/\pi$ from $C_{g,sphere}$ \cite{Purcell85}. One might think that a useful estimate is only obtained by a geometry more similar to the qubit; but the capacitance of an isolated loop of diameter $D$ and wire width $a$ with $D \gg a$ is \cite{Hernandes03}:
\begin{equation}
C_{g,toroid}=\frac{2\pi^2\epsilon_0D}{\mathrm{log}(8D/a)}.
\label{eq:C_toroid}
\end{equation}
For typical loop dimensions and wire widths the logarithmic term is in the range of $5-10$. Therefore, $C_{g,toroid} \approx C_{g,sphere}/3$ which means that the capacitance of a loop is within a factor of three of the capacitance of a sphere with the same linear dimension.

When the loop is on top of a dielectric substrate, $C_{g,toroid}$ is modified. It can be approximated by taking the arithmetic mean of the capacitance when the object is in free space and when it is surrounded by the dielectric, giving
\begin{equation}
C^{\epsilon}_{g,toroid}\approx \frac{\pi^2(\epsilon_{subs}+\epsilon_0)D}{\mathrm{log}(8D/a)}.
\label{eq:C_toroid_sub}
\end{equation}
For typical loop geometries of $D \sim 10-100$ $\mu m$ and $\epsilon_{subs}=10\epsilon_0$ we obtain $C_g \sim 10$ fF.  This is consistent with full, numerical capacitance calculations. While 10fF may seem to be a small capacitance, we will see that it is not negligible and can in fact open the door to significant relaxation. Note also that Eq. \ref{eq:C_toroid_sub} is a lower bound on the capacitance, because it is computed assuming ground is at infinity. In practice, ground is not so distant from the qubit, so that $C_g$ is always somewhat larger than predicted by Eq. \ref{eq:C_toroid_sub}.

The simplest method to quantitatively compute the relaxation time of a qubit is to model it as an LC resonator. This approach has been employed extensively to predict qubit coherence times \cite{Neeley08,Martinis03,Houck08} and we restate the arguments for this approach here for completeness. Note that although the classical model does not, of course, predict the anharmonicity of the qubit, it does accurately predict dissipation. We shall describe the modeling of a simple RF SQUID \cite{Friedman00}. The RF SQUID consists of a Josephson junction embedded in a superconducting loop with inductance $L$ as shown in Fig. 1a. When biased with a flux $\Phi_0/2$, where $\Phi_0=h/2e$ is the flux quantum, the potential is symmetric. Furthermore, the Josephson junction has a phase difference of $\pi$ so that its inductance is approximately equal to $L_J\approx -\Phi_0 / 2 \pi I_0$ where $I_0$ is the critical current of the junction. In order to be a useful flux qubit, the negative Josephson inductance $L_J$ should have a value such that it roughly cancels out the loop inductance: $L_J+L \lsim 0$. From a circuit element perspective, the qubit can now be modeled as two inductors ($L$ and $L_J$) and a capacitance $C$ (the junction self-capacitance or shunting capacitance) all in parallel as shown in Fig. 1b. Suppose the qubit is capacitively grounded ($C_g$) and also capacitively coupled ($C_c$) to a bias lead which, because it is connected to a long transmission line, has an impedance of $Z_0=50$ $\Omega$ as shown in Fig. 1c. The capacitance to ground and the bias lead can be lumped into a single capacitance $C_{eff}=(1/C_g+1/C_c)^{-1}$.

\begin{figure}[t!]
\centering
\includegraphics[width=0.5\textwidth]{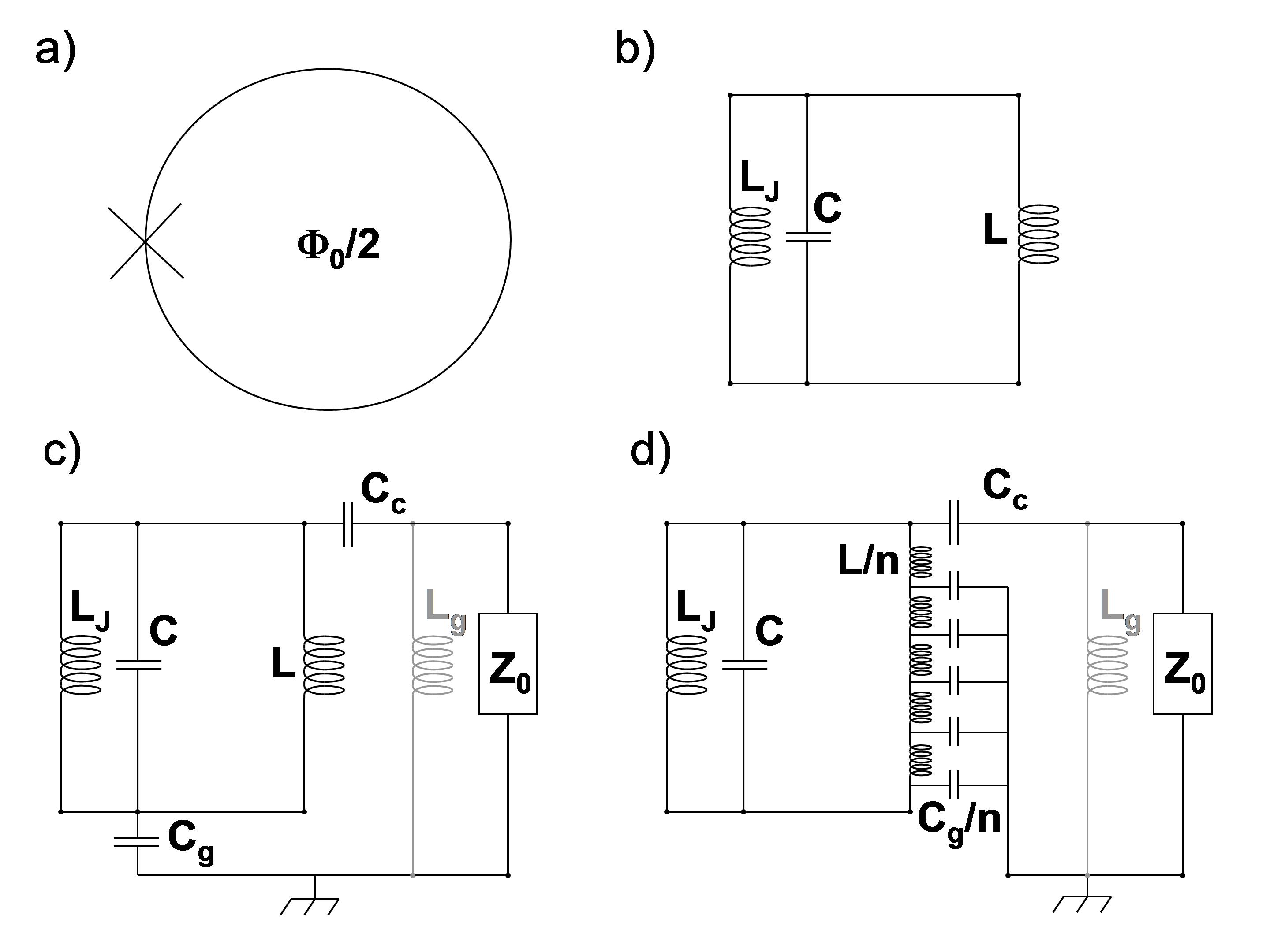}
\label{fig:fig1}
\caption{Sketch of a simple flux qubit and its description using circuit elements. (a) A simple flux qubit consists of a superconducting loop interrupted by a Josephson junction. (b) It can be modeled as a Josephson inductance $L_J$ in parallel with a loop inductance $L$ and junction self-capacitance $C$. (c) Circuit description of the simplest model to capture the qubit's capacitance to ground $C_g$ and to a bias lead $C_c$. (d) A more accurate circuit model discretizes the loop into $n$ segments each with inductance $L/n$ and capacitance to ground $C_g/n$ ($n=5$ is shown).}
\end{figure}

The relaxation time of the qubit at low temperature can now be calculated knowing only the values of the classical circuit elements \cite{Martinis03}. For the RF SQUID, the RLC model tells us that the $T_1$ time is given by a classical $RC$ time constant 
\begin{equation}
T_1=C/Re\{Y\}=CR_{eff}.
\label{eq:t1eq1}
\end{equation}
Here
\begin{equation}
Y=Z_0(\omega C_{eff})^2+i\omega C_{eff}\label{yy}
\end{equation}
is the reactance looking out from the qubit  in the limit $1/\omega C_{ceff} \gg Z_0$, and $R_{eff}=1/Re\{Y\}$ is the effective resistance seen by the qubit.

It is interesting to compare with an apparently very different formula obtained from a standard quantum mechanical treatment (eg. \cite{Martinis03,Burkard04}):
\begin{equation}
T_1=\left( \frac{2\pi}{\Phi_0} \right)^2 \frac{\hbar}{2\omega} \frac{\coth(\hbar\omega/2k_BT)}{|\bra{0}\delta\ket{1}|^2 Re\{Y\}}.\label{eq:qm}
\end{equation}
This formula involves the same reactance $Y$, but also involves a quantum mechanical matrix element of the superconducting phase operator $\delta$ (eg. \cite{Martinis03}).  But for a harmonic system this matrix element can be calculated, with the result $\bra{0}\delta\ket{1}=2\pi/\Phi_0\sqrt{\hbar/2\omega C}$.  With this substitution, the quantum and classical formulas Eqs. (\ref{eq:qm}) and (\ref{eq:t1eq1}) agree exactly in the low temperature limit, where the hyperbolic cotangent factor is one.  

Within the quantum mechanical calculation, we can investigate the change of the matrix element resulting from the small anharmonicity of the qubit potential.  The small resulting rescaling of $T_1$ can be represented in the classical formula by writing
\begin{equation}
T_1=\alpha CR_{eff},
\end{equation} 
with the multiplicative factor $\alpha$ in the range $1<\alpha<3$ for the parameters of a realistic flux qubit.  We will retain this factor in otherwise classical formulas that we discuss below.

Returning to the RF SQUID analysis, with the expression for $Y$ in Eq. (\ref{yy}), we find, in the limit $1/\omega C_{ceff} \gg Z_0$, $R_{eff}=1/Z_0(\omega C_{eff})^2$. The imaginary part of $Y$ is equivalent to a capacitor $C_{eff}$ in parallel with $C$, raising the effective total capacitance of the LC resonator to $C+C_{eff}$ (see ref \cite{Oconnell08}). Therefore, for capacitive coupling to bias leads, and similar to \cite{Oconnell08,Houck08}, we find a relaxation time
\begin{equation}
T^{C_{eff}}_{1} \approx \frac{\alpha(C+C_{eff})}{Z_0(\omega C_{eff})^2}
\label{eq:t1eq2}
\end{equation}
For flux qubits $C \sim 10$ fF. Assuming $\omega/2\pi=5$ GHz, $\alpha=1$, and $C_c = C_g \sim 10$ fF ($C_{eff} \sim 5$ fF) one computes 
\begin{equation}
T_1\approx12ns,
\end{equation}
a very short coherence time compared to the best published results, clearly indicating that capacitive coupling can have a severe impact on coherence times. Note that $C_g \sim 10$ fF is present for typical loop sizes of about $50 \mu m$ as pointed out earlier, and similarly $C_c \sim 10$ fF is easily present in bias loops and/or measurement SQUIDs, particularly because capacitance only scales logarithmically with distance. We next show that coherence times are not significantly altered even when including a more realistic treatment of features such as the distributed nature of the capacitance to ground.

The fact that this capacitance to ground is distributed can be modeled numerically by computing the relaxation time of the circuit of Fig. 1d in which the distributed ground is discretely approximated using $n$ segments. We find that a distributed ground leads to an increase in coherence times by about a factor of $2\leq \beta \leq 5$ for a wide frequency range and a variety of $10\geq C_g/C_c\geq 1$ ratios. Therefore, when $C_g$ is distributed we can write the coherence times as
\begin{equation}
T^{C_{eff,dist}}_1 \approx \frac{\alpha\beta(C+C_{eff})}{50(\omega C_{eff})^2},
\label{eq:t1eq_dist}
\end{equation}
and significant decoherence is present even for a distributed $C_g$.

\section{Symmetry Considerations}
Thus far we have shown that capacitive coupling to bias leads can give rise to short ($\sim 10$ ns) coherence times even for relatively small values for $C_{eff}$. Using our results, in order to obtain coherence times on the order of $1$ $\mu s$ we require $C_{eff}<1$ fF. In order for $C_{eff}$ to be this small, we either require qubit dimensions $D\sim5$ $\mu m$ to obtain $C_g<1$ fF, or alternatively we must make $C_c<1$fF. This result implies that making large dimension flux qubits with long $T_1$ is not possible because $C_g$ cannot be made small, and obtaining less than $1$ fF stray capacitances $C_c$ is exceedingly difficult. Yet, a large flux qubit with coherence times of about $1$ $\mu s$ has been demonstrated \cite{Hime06}, apparently contradicting what we have described thus far. The discrepancy can be explained by invoking symmetry.

Suppose that, with respect to the location of the Josephson junction, the qubit is symmetrically coupled to a bias lead as sketched in Fig. 2a. We find that the relaxation times for such a circuit are infinite (ignoring all other sources of dissipation). A qualitative argument is that no net circulating current can be generated in the qubit loop, similar to arguments for decoherence due to magnetic coupling \cite{Robertson05}. Suppose a voltage source was connected to the coupling capacitor $C_{c1}$ and $C_{c2}$ was absent. In this case a circulating current can be generated. Now suppose only $C_{c2}$ was present. In this case a circulating current can also be generated but the circulation is in the opposite direction. Therefore, when $C_{c1}=C_{c2}$ are both present the net effect is zero. The LC resonator cannot be excited and by reciprocity cannot loose energy \cite{Neeley08}.

\begin{figure}[t]
\centering
\includegraphics[width=0.5\textwidth]{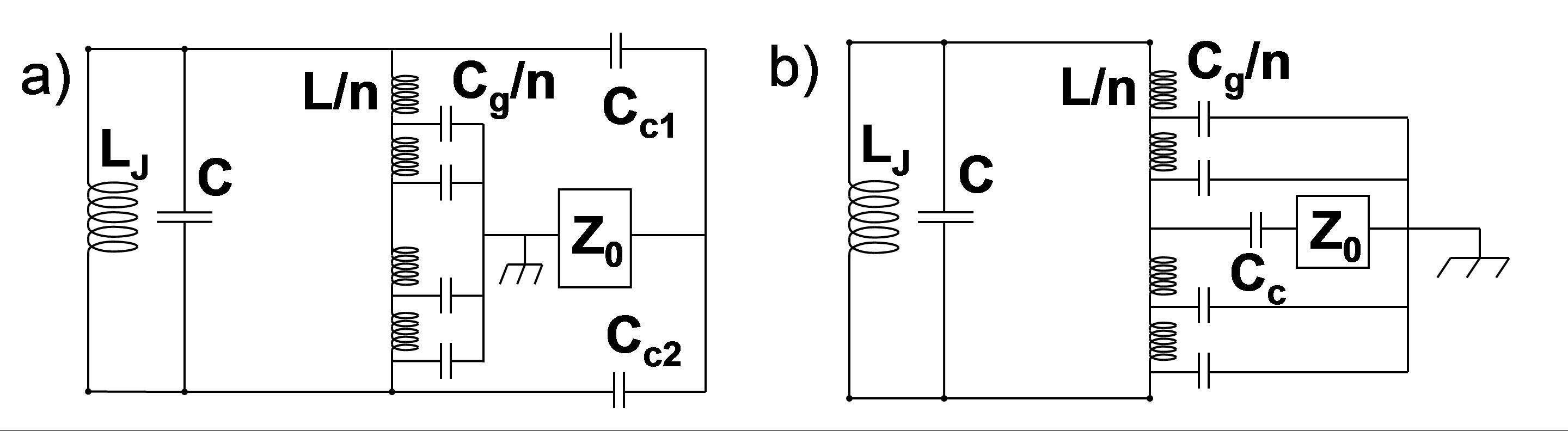}
\label{fig:fig2}
\vspace{-0.25in}
\caption{Examples of circuits which are not impacted by capacitive coupling (shown for $n=4$ - arguments valid for $n \rightarrow \infty$). (a) Coupling to one bias lead symmetrically with capacitances $C_{c1}$ and $C_{c2}$ external noise cannot excite the resonator and therefore it has an infinitely sharp resonance. (b) Coupling to a bias lead to the center of the loop, the circuit also has no loss.}
\end{figure}

Extending this idea of symmetry one can quickly derive that capacitively coupling to the center of the main loop inductance $L$ also gives infinitely long coherence times (see Fig. 2b). This can be seen from Fig. 2a by moving $C_{c1}$ and $C_{c2}$ closer to each other while maintaining symmetry. Eventually $C_{c1}$ and $C_{c2}$ meet in the middle of the loop.

Several qubit results published in the literature are consistent with the observations made here. Smaller flux qubits have better coherence times because of a smaller $C_g$ but more importantly because of symmetry with respect to the junctions. A small, symmetric flux qubit \cite{Chiorescu03} has been show to have long coherence times. However, a slightly larger and asymmetric design \cite{Plantenberg07} has significantly shortened coherence times. A large but highly symmetric flux qubit design has been shown to have good coherence times \cite{Hime06}. This is in contrast with a large flux qubit \cite{Steffen07unpub} with large $C_g$ and $C_c$ which has very short coherence times, even shorter than those reported in ref \cite{Plantenberg07}.

\section{Other Strategies}
Next, we discuss additional methods for reducing the impact of capacitive coupling. By engineering the effective reactance seen by the qubit \cite{Robertson05} it is possible to obtain long coherence times even in the presence of large parasitic capacitances to bias leads. This can be achieved in two possible ways.

The first method is to ground the bias lines, which must be done for flux bias lines and measurement SQUID lines anyway. A sample scenario is shown in Fig. 1c where the qubit is capacitively coupled to the bias line (SQUID or flux bias) which in turn is connected to ground by an inductor $L_g$. If the bias line is well grounded then the capacitive coupling should vanish because no voltage can appear on the bias coil. The effective resistance of this bias circuit is plotted in Fig. 3 for several values of $L_g$ and $C_{eff}=5$ fF. It becomes clear that small inductances to ground are desirable. However, what is surprising is that even an inductance of $1$ nH gives an effective resistance that is only about $2-10$ times larger than if the inductance to ground was infinite. This value of inductance is present for wires that are only $1$ mm in length and therefore as far as decoherence is concerned a connection to ground of more than $1$ mm is as poor as one that is absent or infinitely long. In order to achieve significant gains in the effective resistance (about 2-3 orders of magnitude), the inductance to ground should be made $100$ pH corresponding to a wire length of only $100$ $\mu m$.

\begin{figure}[t!]
\centering
\includegraphics[width=0.5\textwidth]{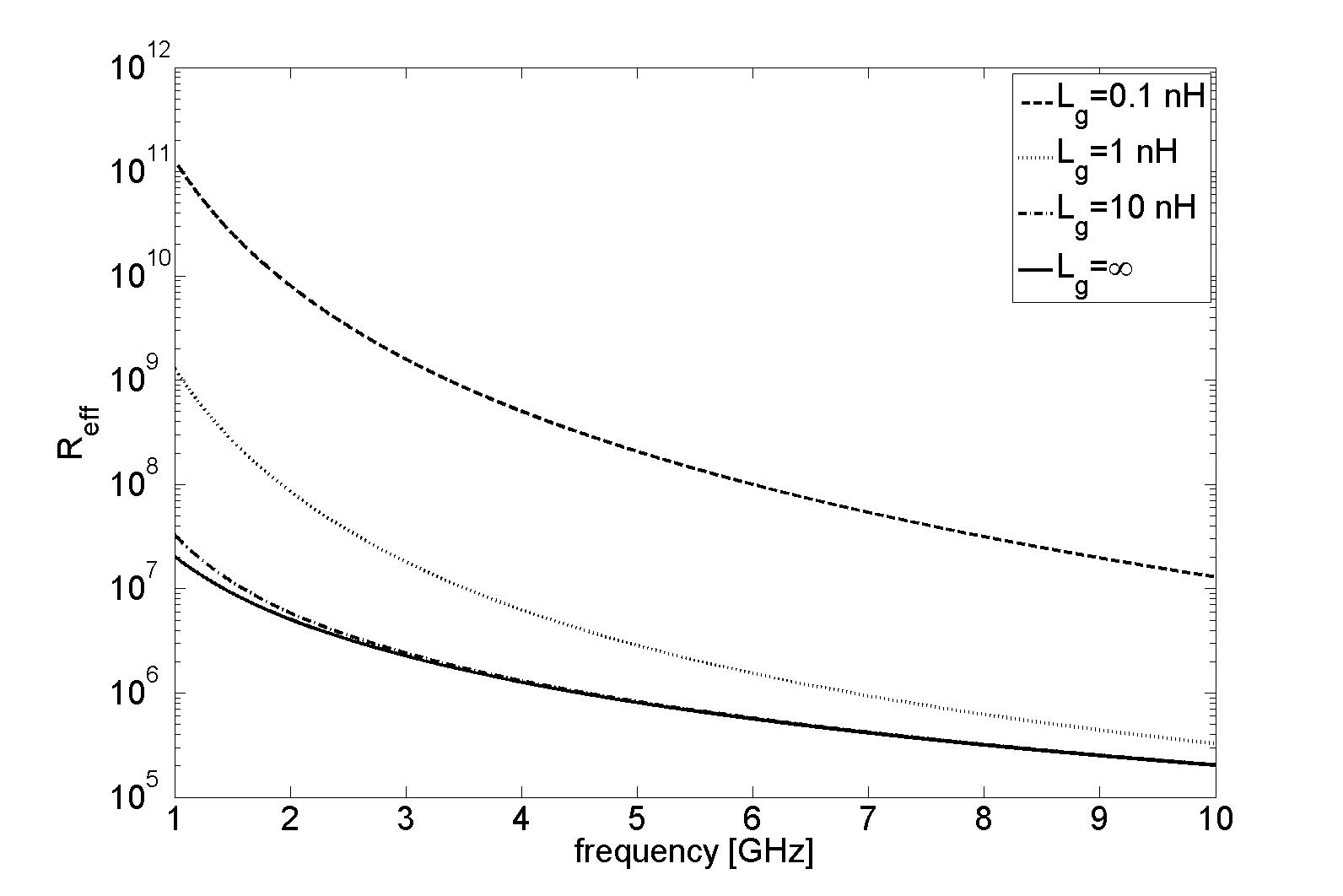}
\label{fig:fig3}
\vspace{-0.25in}
\caption{Transformed impedance when capacitively coupling to an inductively grounded bias line ($C_{eff}=5$ fF).}
\end{figure}

The second method includes inserting choke inductors or capacitors to ground into the bias line in front of the coupling capacitance to improve the effective reactance even further \cite{Robertson05}. The exact values for inductors and capacitors depend on the required bandwidth of the bias lines which is typically less than $1$ GHz except for microwave lines. It is therefore conceivable to engineer a large effective resistance at qubit frequencies greater than $5-6$ GHz.

\section{Discussion}
Finally, we shall make some qualitative remarks about $T_2$ dephasing times within the setting of capacitive coupling. Besides the limit to dephasing from $T_1$ times ($T_2=2T_1$) we believe there are no significant contributions to dephasing from capacitively coupling to bias leads, in particular for flux qubits. Dephasing is the result of low frequency noise that leads to modulations of the qubit resonance frequency. Because we are concerned with capacitive coupling low frequency noise should not easily couple into the qubit. As a result, the integrated noise should be small. Additionally, even if some low frequency noise reaches the qubit, leading to asymmetric current flow, the flux qubit should retain long dephasing times because its resonance frequency to first order does not vary with the bias flux when biased at $\Phi_0/2$. Dissipation should therefore remain as the most significant source of decoherence.

In summary, we have shown that it is hard to prevent the capacitance to ground from being a significant or even dominant contributor to decoherence for floating qubits. The classical formulas for relaxation times as well as the dependence of these on symmetry set a hard limit on the degree of quantum coherence that is possible in these systems. While various parts of our arguments (capacitance to ground, dissipation from capacitive coupling to bias leads in resonators, symmetry) have been touched on in the literature, they have not been previously been combined to give a full picture of the expected capacitive losses in floating qubits. While our estimates for the degree of anharmonicity and the amount of distributed vs. lumped capacitance will not apply to all experiments, we believe that the qualitative aspects of our predictions will be very widely applicable. Because we predict potentially short coherence times in asymmetric floating qubit designs and because an eventual quantum computer requires very long coherence times, it is clear that careful attention must be paid to the impact that capacitive coupling has on the prospects of scalability. We are currently concerned about the prospects of scalability for qubits with small self-capacitance: since $T_1 \propto C$, even small asymmetries can lead to a drastic reduction in coherence times. In addition to qubit-qubit interactions due to capacitive coupling, it is also not clear how to arrange multiple qubits and their associated inputs/outputs in a symmetric fashion to minimize capacitive coupling. On the other hand, qubit designs with much larger self-capacitances exist (e.g. phase qubits). Although capacitive-coupling dissipation will certainly also occur in these systems, they may occur at a more manageable level. 

The authors would like to thank stimulating discussions with John Clarke, John Martinis and Robert McDermott.

\bibliography{steffen_capacitance_Re_NJP}

\begin{thebibliography}{24}
\bibitem[{com()}]{commentc}
\bibinfo{note}{We have also used the fully quantum mechanical models to confirm
  the symmetry arguments.}
\end{thebibliography}

\end{document}